\documentclass[amsmath, aps]{revtex4}
\usepackage{graphicx}
\usepackage{times}
\usepackage{epsfig}

\begin{document}
\title{Two-photon decay of excited levels in hydrogen: the ambiguity of the separation 
of cascades and pure two-photon emission}

\author{L. Labzowsky$^{1),2)}$, D. Solovyev$^{1)}$ and G. Plunien$^{3)}$}

\affiliation{ 1) V. A. Fock Institute of Physics, St. Petersburg
State University, Petrodvorets, Oulianovskaya 1, 198504, St.
Petersburg, Russia
\\
2) Petersburg Nuclear Physics Institute, 188300, Gatchina, St.
Petersburg, Russia
\\
3) Technische Universit\"{a}t Dresden, Mommsenstrasse 13, D-01062,
Dresden, Germany.}

\begin{abstract}
The problem of the evaluation of the two-photon decay width of excited states in hydrogen is considered. Two different approaches to the evaluation of the width including cascades channels are employed: the summation of the transition probabilities for various decay channels and the evaluation of the imaginary part of the Lamb shift. As application the decay channels for the $3s$ level of the hydrogen atom are evaluated, including the cascade transition probability $3s-2p-1s$ as well as "pure" two-photon decay probability and the interference between both channels. 
An important role should be assigned to the "pure" two-photon probability in astrophysics context,
since processes of this kind provide a possibility for the decoupling of radiation and 
matter in the early Universe. 
We demonstrate the ambiguity of separation of the "pure" two-photon contribution and criticize the existing attempts for such a separation.
\end{abstract}
\maketitle

\section{Introduction}
During recent years the two-photon decay processes in hydrogen have attracted special attention due to new and very accurate observations of the cosmic microwave background temperature and polarization anisotropies \cite{Hinshaw,Page}. 
In view of these observations it becomes important to understand the hydrogen recombination history with high precision. In the early Universe the strong Lyman-alpha $2p-1s$ transition did not permit the atoms to remain in their ground states: each photon released in such a transition in 
one atom was immediately absorbed by another one. However, due to the very weak $2s-1s$ two-photon decay process the radiation could finally decouple from the interaction with matter and thus 
permit a successful recombination. The role of the $2s-1s$ two-photon decay was first established in \cite{Zeldovich, Peebles}. The other two-photon channels, i.e. $ns-1s$, $nd-1s$ decays were also discussed in \cite{cea86}-\cite{Wong}. At the present level of accuracy reached in the astrophysical observations these contributions also become important.

A crucial difference between the decay of the $ns$ (with $n>2$, similar for $nd$) 
and the $2s$ levels consists in the presence of the cascade transitions as the dominant decay channels. In case of the $2s$ level the cascade transitions are absent. Since the cascade photons can be  effectively reabsorbed, the problem of separating the "pure" two-photon contribution from the cascade contribution arises. An interference between the two decay channels, i.e. "pure" two-photon decay and cascade, should also be taken into account.

A similar problem did arise within the theory of the two-electron highly charged ions (HCI) \cite{Drake}-\cite{LabShon}. Drake \cite{Drake} evaluated for the first time the 
two-photon E1M1 transition in the presence of a cascade transition in He-like uranium ion ($Z=92$). Later Savukov and Johnson preformed similar calculation for a variety of  He-like HCI ($50\leq Z\leq 92$) \cite{Savukov}. In \cite{Drake, Savukov} the "pure" two-photon contribution was obtained by subtracting off a Lorentzian fit for the cascade contribution from the total two-photon decay frequency distribution. The existence of the interference terms was recognized in \cite{Drake, Savukov}, but only approximately included in the Lorentzian fit as an asymmetric deviation from the pure Lorentzian. A rigorous QED approach for the evaluation of two-photon decay width with cascades was developed in \cite{LabShon} (see also \cite{AndrLab}). This approach was based on the standard evaluation of the decay probability as a transition probability to the lower levels. In case of cascades the integral over emitted photon frequency distribution becomes divergent due to the singular terms, corresponding to the cascade resonances. To avoid such a singularity, the 
resummation of an infinite series of the electron self-energy insertions was performed in
\cite{LabShon}. This resummation converts into a geometric progression and in this way the self-energy insertion (and the level width as its imaginary part) enters in the energy denominator and shifts the pole on the real axis into the complex energy plane, thus making the integral finite. With this approach F. Low first derived the Lorentz profile from QED \cite{Low}. In \cite{Drake, Savukov} the level widths in the singular energy denominators were also introduced, though without special justification. Similarly, i.e. by introducion of the level widths in the singular energy denominators the two-photon decay of $ns$ and $nd$ excited levels in hydrogen was evaluated 
most recently in the astrophysical papers \cite{Chluba, Hirata}.

In \cite{LabShon} the ambiguity of the separation of the "pure" two-photon decay and cascades was first revealed for HCI, where it was shown that the interference terms can essentially contribute 
to the total decay probability. The full QED treatment of the two-photon decay
process with inclusion of cascades and interference terms was performed in
\cite{Chluba, Hirata} for the hydrogen atom. However, in \cite{Chluba, Hirata}
the ambiguity of separation of the "pure" two-photon decay was neither
emphasized nor demonstrated explicitly. This will be subject of the present
paper. Our numerical results for $3s$-level in hydrogen are in agreement with
the most accurate recent calculations in \cite{Chluba} (see Section 11
below). However, our results disagree strongly with the value obtained in
\cite{jas08}. The result in \cite{jas08} follows from the "alternative"
approach to the evaluation of the two-photon decay, developed by Jentschura
\cite{Jent1, Jent2, Jent3}. This "alternative" approach is based on the evaluation of the imaginary part of the two-loop Lanb shift for $3s$-level. The idea to use the imaginary part of the Lamb shift for calculating the radiative corrections to the one-photon transitions is due to Barbieri and Sucher \cite{bas78} and is definitely adequate for this propose. Later this approach was used in \cite{Sapir} for HCI.
Still, the extension of this approach to the evaluation of the two-photon decay width, when the cascade transitions are also present, is more involved and requires special care. 
In \cite{Jent1, Jent2, Jent3} it was claimed that the singularity in the integration over the photon frequency distribution for the two-photon decay is absent and thus the finite integral represents directly the "pure" two-photon decay rate.

Also this statement forced us to perform a careful revision of the evaluation of the two-photon decay width via the imaginary part of the second order Lamb shift, which will be presented below in Sections 6-10. In our approach we employed the consequences of the "optical theorem" for the $S$-matrix and the Gell-Mann and Low adiabatic formula \cite{Gell} for the evaluation of the energy level shift via $S$-matrix. The results of our analysis, unlike Jentschura's derivations, reveal
that the evaluation of the two-photon decay width via the imaginary part of the Lamb shift provides identical expressions as the standard QED description via summation of transition probabilities. 
The integration over the emitted photon frequency for the two-photon decay with cascades remains divergent and requires the introduction of the level widths in the singular energy denominators
(resummation of QED-radiative corrections). 
The detailed derivations employing both methods (summation of transition probabilities and 
evaluation of the imaginary part of the Lamb shift) are given and numerical results for the 
$3s$-level in hydrogen are provided. These results demonstrate clearly the ambiguity of the separation of the "pure" two-photon and cascade contributions with accuracy, required for the modern astrophysical investigations (i.e. 1\%).

Our paper is organized as follows. In Section 2 we shortly review the standard derivation of the excited atomic level one-photon width via the summation over transition probabilities to the lower levels. This presentation of formulas, though available in text-books on quantum electrodynamics, are necessary for the further comparison with the results obtained by another approaches. 
It is for the readers convenience to make the paper self-contained and allowing to follow our derivations without consulting additioanl literature. The same is done (even in a more compressed way) in the Section III for the two-photon decay width. In Section 4 we repeat (again very briefly) the standard understanding of the situation for the two-photon decay with cascades and present the formulas employed for the appropriate calculations in the numerous works on the subject, beginning from \cite{Drake}. For justification of this procedure we follow \cite{LabShon}. In Section 5 we evaluate the one-photon decay width via the imaginary part of the Lamb shift. This is again a well-known result but it will be also necessary to confirm our further derivations. In this Section we follow derivations of  \cite{LabKlim} in a more condensed way.

In Section VI we evaluate the imaginary part of the energy level shift employing the adiabatic formula of Gell-Mann and Low. The key idea of our approach - the application of the "optical theorem" - is formulated in Section VII. Actually, we employ here the unitarity of the adiabatic $S$-matrix. The formulated approach is applied to the evaluation of the one-photon width in the Section VIII, to the evaluation of the two-photon width (without cascades) in Section IX and to the evaluation of the two-photon width (with cascades) in Section X. In these sections we compare our derivations with the standard QED derivations and analyse the discrepance with Jentschura's results in Section X. In Section XI the numerical results for the $3s$-level in hydrogen are given and the ambiguity of separation the "pure" two-photon and cascade contributions is demonstrated. Finally, in Section XII we formulate our version of the problem of the cascade separation in astrophysics and present our recommendations for the resolution of this problem.
Relativistic units $\hbar=c=1$ are employed.

\section{One-photon decay width via summation of transition probabilities}
\subsection{Transition probability}
The first-oder matrix element of the $S$-matrix describing the one-photon emission in a 
one-electron atom (see Fig. 1) is given by 
\begin{eqnarray}
\label{1}
\langle A'|\hat{S}^{(1)}|A\rangle  = e \int d^4 x\, \bar{\psi}_{A'}(x)\gamma_{\mu}A^*_{\mu}(x)\psi_A(x)\,.
\end{eqnarray}
Here $\hat{S}^{(1)}$ is the first-order $S$-matrix, $e$ is the electron charge, $\psi_A(x) = \psi_A(\vec{r})e^{-i E_A t}$, $\psi_A(\vec{r})$ is the solution of the Dirac equation for the atomic electron, $E_A$ is the Dirac energy, $\bar{\psi}_{A'} = \psi_{A'}^\dagger \gamma_0$ is the Dirac conjugated wave function with $\psi_{A'}^{\dagger}$ being its Hermitian conjugate and 
$\gamma_{\mu} = (\gamma_0, \vec\gamma)$ are the Dirac matrices. 
The photon field $A_{\mu}(x)$ appears in terms of eigenmodes of the form 
\begin{eqnarray}
\label{2}
A^{(\vec e,\,\vec k)}_{\mu}(x) = \sqrt{\frac{2\pi}{\omega}}\,e^{(\lambda)}_{\mu}e^{i(\vec{k}\vec{r}-\omega t)}
 = \sqrt{\frac{2\pi}{\omega}}\,e^{(\lambda)}_{\mu}e^{-i\omega t}\,A^{(\vec e,\,\vec k)}_{\mu}(\vec r\,)
\, ,
\end{eqnarray}
where $e^{(\lambda)}_{\mu}$ is the photon polarization 4-vector, $k=(\vec{k},\omega)$ is the photon momentum 4-vector ($\vec{k}$ is the wave vector, $\omega=|\vec{k}|$ is the photon frequency), $x\equiv(\vec{r},t)$ is the coordinate 4-vector ($\vec{r}, t$ are the space- and time-coordinates). 

Describing emitted (real) photons in the Coulomb gauge implies the transversality condition 
\begin{eqnarray}
\label{3}
\gamma_{\mu}e^{(\lambda)}_{\mu}=   \vec{e}^{\,(\lambda)} \vec{\gamma}\, ,
\end{eqnarray}
where $  \vec{e}^{\,(\lambda)}$ is the 3-vector of the photon polarization. Then integrating over the time variable yields
\begin{eqnarray}
\label{4}
\langle A'|\hat{S}^{(1)}|A\rangle &=&  2\pi\,\left(\vec{A}^*_{\vec e, \vec k} \vec{\alpha}\right)_{A'A}\,
\delta\left(\omega-E_A+E_{A'}\right) \nonumber \\
&=& 2\pi\,e\sqrt{\frac{2\pi}{\omega}}\,\left(( \vec{e}^{\,*}\vec{\alpha})\,
e^{-i\vec{k}\vec{r}}\right)_{A'A}\,\delta\left(\omega-E_A+E_{A'}\right)\, .
\end{eqnarray}
Here $\vec{\alpha} = \gamma_0\vec\gamma$ are the Dirac matrices and 
$(\dots )_{A'A}$ denotes the spatial matrix element $\langle A'|\dots |A\rangle$ with the wave functions  $\psi^{\dagger}_{A'}(\vec{r}) = \langle A'|\vec r\rangle$ and 
$\psi_{A}(\vec{r}) = \langle\vec r|A\rangle$, respectively. 

The transition amplitude $U_{A'A}$ is defined as 
\begin{eqnarray}
\label{5}
\langle A'|\hat{S}^{(1)}|A\rangle = -2\pi\, i\delta\left(\omega-E_A+E_{A'}\right)U_{A'A}^{(1)}\, .
\end{eqnarray}

The transition probability cannot be defined straightforeward as $\left|\langle A'|\hat{S}^{(1)}|A\rangle \right|^2$, since the square of a $\delta$-function is actually meaningless. 
The standard way to overcome this technical diffculty is to express one of the two 
$\delta$-function, which arises in Eq. (\ref{1}) after integration over times $t$ 
as the Fourier integral
\begin{eqnarray}
\label{6}
\delta(E)=\frac{1}{2\pi}\int\limits_{-\infty}^{\infty}e^{iEt}dt\, ,
\end{eqnarray}
by a representation $\delta_T(E)$. The latter is introduced by restricting the integration 
to the finite time interval $(-T/2,+T/2)$ with the result 
\begin{eqnarray}
\label{7}
\delta_T(E) = \frac{1}{2\pi}\int\limits_{-T/2}^{T/2}e^{iEt}dt\, .
\end{eqnarray}
Multiplying $\delta_T(E)$ by the second $\delta$-function $\delta(E)$ results in the substitution
\begin{eqnarray}
\nonumber
\delta (E) \delta_T(E) =  \delta (E)\delta_T(0) =\delta (E) \frac{T}{2\pi}\, .
\end{eqnarray}
Thus, the probability of the process appears to be proportional to the observation time interval $T$. It is natural then to introduce the transition probability per time unit (transition rate) by definition
\begin{eqnarray}
\label{8}
W_{A'A}=\lim_{T\rightarrow\infty}\,\frac{1}{T}\left|\langle A'|\hat{S}^{(1)}(T)|A\rangle\right|^2
= 2\pi\left|U_{A'A}^{(1)}\right|^2\delta\left(\omega-E_A+E_{A'}\right)\, .
\end{eqnarray}
We will compare this definition with another one in Section 8.

If the final state belongs to the continuous spectrum (due to the emitted photon in our case) the differential transition probability should be introduced
\begin{eqnarray}
\label{9}
dW_{A'A}(\vec{k},\vec{e})=2\pi\left|U_{A'A}^{(1)}\right|^2\delta\left(\omega-E_A+E_{A'}\right)
\frac{d\vec k}{(2\pi)^3}\, ,
\end{eqnarray}
where $d\vec k\equiv d^3k = d\Omega_{\vec\nu}d\omega \omega^2$.
Integration in Eq. (\ref{9}) over $\omega$ gives the probability of the photon emission with polarization $\vec{e}$ in the direction $\vec{\nu}\equiv\vec{k}/\omega$ per time unit (and solid angle
$d\Omega_{\vec\nu}\equiv d\vec\nu$):
\begin{eqnarray}\label{10}
dW_{A'A}=\frac{e^2}{2\pi}\omega_{A'A}\left|\left(( \vec{e}^{\,*}\vec{\alpha})e^{-i\vec{k}\vec{r}}\right)_{A'A}\right|^2d\vec{\nu}\, ,
\end{eqnarray}
where $\omega_{A'A}=E_A-E_{A'}$. Note, that due to the presence of $\delta$-function formally we should integrate over the frequency interval $(-\infty, \infty)$, while physically the frequency of real photons changes within the interval $(0, \infty)$. 
The total transition probability follows from Eq. (\ref{10}) after integration over angles and summation over the polarization:
\begin{eqnarray}
\label{11}
W_{A'A}=\frac{e^2}{2\pi}\omega_{A'A}\sum\limits_{\vec{e}}\int d\vec{\nu}\left|\left(( \vec{e}^{\,*}\vec{\alpha})e^{-i\vec{k}\vec{r}}\right)_{A'A}\right|^2
\end{eqnarray}
The result of this summation and integration will be carried out in Section 5.

\subsection{Nonrelativistic limit}

For the atomic electron the characteristic scales for $|\vec{r}|$ and $|\vec{k}|=\omega$ are: $|\vec{r}|\sim 1/m\alpha Z$, $\omega=E_{A'}-E_A\sim m(\alpha Z)^2$, where $m$ is the electron mass, $\alpha$ is the fine structure constant, $Z$ is the charge of the nucleus. Then in the nonrelativistic case, in particular for 
the hydrogen atom ($Z=1$), the exponential function
in the matrix element in Eq. (\ref{11}) can be replaced by 1, since $\vec{k}\vec{r}\sim \alpha$. 
In the nonrelativistic limit the matrix element involving the Dirac matrices $\vec{\alpha}$ 
(electron velocity operator $\hat{\vec{v}}$ in the relativistc theory) 
can be substituted by $\hat{\vec{p}}/m$, where $\hat{\vec{p}}$ is the electron momentum operator. 
Then Eq. (\ref{11}) takes the form 
\begin{eqnarray}
\label{12}
W_{A'A}=\frac{e^2}{2\pi m^2}\omega_{A'A}\sum\limits_{\vec{e}}\int d\vec{\nu}\left|\left(\vec{e}\vec{p}\right)_{A'A}\right|^2\, ,
\end{eqnarray}
where the notation $(...)_{A'A}$ now also implies evaluation of the matrix element with Schr\"{o}dinger 
wave functions.

Performing summation over the polarization with the help of the known formula
\begin{eqnarray}
\label{13}
F=\sum\limits_{\vec{e}}( \vec{e}^{\,*}\vec{a})(\vec{e}\vec{b})=(\vec{a}\times\vec{\nu})(\vec{b}\times{\vec{\nu}})
\end{eqnarray}
with $\vec{a}$, $\vec{b}$ being two arbitrary vectors, we arrive at the standard expression for 
the one-photon probability in the "velocity" form after integrating over $\vec{\nu}$:
\begin{eqnarray}
\label{14}
W_{A'A}=\frac{4}{3}\frac{e^2}{m^2}\omega_{A'A}\left|\left(\vec{p}\right)_{A'A}\right|^2\, .
\end{eqnarray}
The so-called "length" form 
\begin{eqnarray}
\label{16}
W_{A'A}=\frac{4}{3}\omega_{A'A}^3\left|(\vec{d})_{A'A}\right|^2
\end{eqnarray}
involving the electric dipole moment operator $\vec{d}=e\vec{r}$ of the electron
can be obtained via the quantum mechanical relation
\begin{eqnarray}
\label{15}
\omega_{A'A}(\vec{r})_{A'A}=\frac{i}{m}(\vec{p})_{A'A}\, .
\end{eqnarray}
Eq. (\ref{16}) reveals that in the non-relativistic limit the atomic radiation is essentially
the radiation of the electric dipole.

In the nonrelativistic limit the total one-photon width of the atomic level $A$ 
results as
\begin{eqnarray}
\label{17}
\Gamma_{A}^{(1)}=\frac{4}{3}\omega_{A'A}^3\sum\limits_{E_{A'}<E_A}\left|(\vec{d})_{A'A}\right|^2,
\end{eqnarray}
where the summation runs over all levels $A'$ with energy $E_{A'}$ lower than that of the level 
$A$ (provided the transitions $A\rightarrow A'$ are allowed in the nonrelativistic limit).

\section{Two-photon decay width via summation of transition probabilities}

The two-photon transition probability $A\rightarrow A'+2\gamma$ corresponds to the following second-order $S$-matrix elements (see Fig. 2)
\begin{eqnarray}
\label{18}
\langle A'|\hat{S}^{(2)}|A\rangle  = e^2\int d^4x_1d^4x_2\left(\bar{\psi}_{A'}(x_1)\gamma_{\mu_1}A^*_{\mu_1}(x_1)S(x_1x_2)\gamma_{\mu_2}A^*_{\mu_2}(x_2)\psi_A(x_2)\right),
\end{eqnarray}
where $S(x_1x_2)$ is the Feynman propagator for the atomic electron. 
In the Furry picture the eigenmode decomposition reads (e.g. \cite{Akhiezer})
\begin{eqnarray}
\label{19}
S(x_1x_2)=\frac{1}{2\pi i}\int\limits_{\infty}^{\infty}d\omega_1e^{i\omega_1(t_1-t_2)}\sum\limits_n\frac{\psi_n(x_1)\bar{\psi}_n(x_2)}{E_n(1-i0)+\omega_1}\, ,
\end{eqnarray}
where the summation in Eq. (\ref{19}) extends over the entire Dirac spectrum of electron states $n$ in the field of the nucleus. Using again Eqs. (\ref{5}) and (\ref{9}) for the two-photon transition and integrating over time and frequency variables in Eq. (\ref{18}) for the sum of the contributions of the both Feynamn graphs (see Figs. 2a, 2b),  we find
\begin{eqnarray}
\label{20}
dW_{A'A}=2\pi\delta\left(E_A-E_{A'}-\omega-\omega'\right)\left|U_{A'A}^{(2)}\right|^2\frac{d\vec{k}}{(2\pi)^3}\frac{d\vec{k}'}{(2\pi)^3}\, ,
\end{eqnarray}
\begin{eqnarray}
\label{21}
U_{A'A}^{(2)} = \frac{2\pi e^2}{\sqrt{\omega\omega'}}\left[\sum\limits_n\frac{\left(\vec{\alpha}\vec{A}^*_{\vec{e},\vec{k}}\right)_{A'n}\left(\vec{\alpha}\vec{A}^*_{ \vec{e}\,',\vec{k}'}\right)_{nA}}{E_n-E_A+\omega'}+\sum\limits_n\frac{\left(\vec{\alpha}\vec{A}^*_{ \vec{e}\,',\vec{k}'}\right)_{A'n}\left(\vec{\alpha}\vec{A}^*_{\vec{e},\vec{k}}\right)_{nA}}{E_n-E_A+\omega}\right] 
\end{eqnarray}
with the abbreviatory notation
$\vec{A}_{\vec{e}, \vec{k}} = \vec{e}\,e^{i\vec{k}\vec{r}}$.

In what follows, we will be interested in the decay width of the $ns$ levels 
($A\equiv ns \rightarrow A'\equiv 1s$)
in hydrogen. In this section we focus on the case $n=2$, when the cascades are absent. In the nonrelativistic limit, after the integration over frequencies $\omega'$, over photon directions 
$d\vec{\nu}$, $d\vec{\nu}'$ and summation over all  polarizations $\vec{e}$, $ \vec{e}\,'$ is performed, we obtain for the photon frequency distribution:  
\begin{eqnarray}
\label{23}
dW_{ns,1s}(\omega)=\frac{8\omega^3(\omega_0-\omega)^3}{27\pi}e^4\left|S_{1s,ns}(\omega)+S_{1s,ns}(\omega_0-\omega)\right|^2d\omega\, ,
\end{eqnarray}
\begin{eqnarray}
\label{24}
S_{1s,ns}(\omega)=\sum\limits_{n'p}\frac{\langle R_{1s}|r|R_{n'p}\rangle\langle R_{n'p}|r|R_{ns}\rangle}{E_{n'p}-E_{ns}+\omega}\, ,
\end{eqnarray}
\begin{eqnarray}
\label{25}
\langle R_{n'l'}|r|R_{nl}\rangle=\int\limits_{0}^{\infty}r^3R_{n'l'}(r)R_{nl}(r)dr\, ,
\end{eqnarray}
where $\omega_0=E_{ns}-E_{1s}$, $R_{nl}(r)$ are the radial part of the nonrelativistic hydrogen wave functions, and $E_{nl}$ are the hydrogen electron energies. Here we have used again the quantum-mechanical relation Eq. (\ref{15}); Eq. (\ref{23}) is written in the "length" form.

The decay rate for the two-photon transition can be obtained by integration of Eq. (\ref{1}) over the entire frequency interval 
\begin{eqnarray}
\label{26}
W_{ns,1s}=\frac{1}{2}\int\limits_0^{\omega_0}dW_{ns,1s}(\omega).
\end{eqnarray}
In case $n=2$ the cascade transitions are absent, the frequency distribution Eq. (\ref{23}) is not singular and the integral Eq. (\ref{26}) is convergent.

\section{Two-photon decay with cascades}

In case of the cascade transitions ($n> 2$), some terms in Eq. (\ref{24}) become singular and the integral Eq. (\ref{26}) diverges. This divergency has a physical origin: an emitted photon meets the resonance. So the divergency can be avoided only by introducing the width of this resonance.This situation was studied in \cite{LabShon} for the HCI. The same recipe can also be used in case of the hydrogen atom. Following the prescriptions given in \cite{LabShon} we separate out the resonant terms (corresponding to cascades) in the sum over the intermediate states Eq. (\ref{24}) and apply Low's procedure \cite{Low} for the regularization of the corresponding expressions in the vicinity of the resonance frequency values. Practically this 
leads to the apperance of the energy level widths in the energy denominators. Then the Lorentz profiles arise for the resonant terms in the expression for the probability. However, the Lorentz profile is valid only in the vicinity of the resonance and cannot be extended too far off 
from the resonance frequency value. As for any multichannel processes such a separation is an approximate procedure due to existence of the interference terms.

The integration over the entire frequency interval $[0,\omega_0]$ in Eq. (\ref{26}) should be split into several subintervals, e.g. 5 in case of the two-photon emission profile for the $3s$-level decay, see Fig. 3. The first interval (I) extends from $\omega=0$ up to the lower boundary of the second interval (II). The latter one encloses the resonance frequency value $\omega_1=E_{3s}-E_{2p}$. Within the interval (II) the resonant term $n=2$ in Eq. (\ref{24}) should be subtracted from the sum over intermediate states and replaced by the term with modified energy denominator. This modified denominator is $E_{2p}-E_{3s}+\omega+\frac{i}{2}\Gamma$, where $\Gamma=\Gamma_{2p}+\Gamma_{3s}$. The third interval (III) extends from the upper boundary of interval II up to the lower boundary of the interval (IV), the latter one enclosing another resonance frequency value $\omega_2=E_{2p}-E_{1s}$. Within the interval (IV) again the resonant term $n=2$ in Eq. (\ref{24}) should be replaced by the term with modified denominator $E_{2p}-E_{1s}-\omega-\frac{i}{2}\Gamma_{2p}$. Finally, a fifth interval (V) ranges from the upper boundary of the interval (IV) up to the maximum frequency value $\omega_0$. Note, that the frequency distribution $dW_{3s,1s}(\omega)$ is symmetric with respect to $\omega=\omega_0/2$ with a 1\% accuracy (the asymmetry is due to the difference between $\Gamma=\Gamma_{2p}+\Gamma_{2s}$ and $\Gamma_{2p}$, respectively). The discussion on the choice of the size of the intervals (II) and
(IV), which defines also the size of the other intervals, as well as on further approximations 
will be postponed until Section 11.

In both Section III, IV the total width of the levels $\Gamma_{ns}$ is defined by the sum of the one-photon and two-photon transition rates to the lower levels. The cascade transitions yield the dominant contribution to $\Gamma_{ns}$ except for the case $n=2$.

\section{One-photon decay width via the imaginary part of the Lamb shift: direct evaluation}

In this Section we will briefly recall the well known derivation of the one-photon level width arising from pure one-photon
transitions via the imaginary part of the Lamb shift. This may serve as a lucid introduction to the further derivations with
employment of the adiabatic $S$-matrix theory. 
Let us consider a (free) one-electron ion or atom in the excited state $|A\rangle$ interacting 
with the vacuum $|0_\gamma\rangle$ of the quantized radiation field (with no additional external electromagnetic fields
present). The initial state of the total system ''atom + radiation field'' 
$|I\rangle = |A\rangle|0_\gamma\rangle \equiv |A, 0_\gamma\rangle$ 
is described as pure number eigenstate. 
Since the one-loop vacuum-polarization contribution to the Lamb shift is real, 
the pure one-photon width of the excited level $A$ is given by the imaginary part of the one-loop electron self-energy contribution to the Lamb shift for the level $A$ (see Fig. 4):
\begin{eqnarray}
\label{27}
\Delta E^{(2)}_A = Re\Delta E^{(2)}_A+i Im \Delta E^{(2)}_A = Re \Delta E^{(2)}_A - \frac{i}{2}\Gamma^{(1)}_A.
\end{eqnarray}
Note, that the superscript at $\Delta E^{(2)}$ refers to powers of the coupling constant $e$, while the superscript at $ \Gamma^{(1)}_A$ implies the number of emitted photons.  
The second-order $S$-matrix element, which corresponds to the Fig. 4, looks like 
\begin{eqnarray}
\label{28}
\langle A|S^{(2)}|A\rangle  = \int d^4x_1d^4x_2\left(\bar{\psi}_A(x_1)\gamma_{\mu_1} S(x_1x_2)\gamma_{\mu_2}\psi_A(x_2)\right)D_{\mu_1\mu_2}(x_1x_2)\, ,
\end{eqnarray}
where $D_{\mu_1\mu_2}(x_1x_2)$ is the photon propagator in the Feynman gauge:
\begin{eqnarray}
\label{29}
D_{\mu_1\mu_2}(x_1x_2)=\frac{1}{2\pi i}\frac{\delta_{\mu_1\mu_2}}{r_{12}}\int\limits_{-\infty}^{\infty}d\omega_1e^{i\omega_1(t_1-t_2)+i|\omega_1|r_{12}},
\end{eqnarray}
$r_{12}=|\vec{r}_1-\vec{r}_2|$. For the evaluation of this energy shift corresponding to an "irreducible" Feynman 
graph (i.e. such graphs cannot be cut into subgraphs of lower order by cutting only the electron lines; the graph Fig. 3 belongs to the "irreducible" ones) the following formula can be used:
\begin{eqnarray}
\label{30}
\Delta E^{(2)}_A=\langle A|U^{(2)}|A\rangle _{irr},
\end{eqnarray}
where the amplitude $U^{(2)}$ is defined by the relation
\begin{eqnarray}
\label{31}
\langle A'|S^{(2)}|A\rangle =-2\pi i\delta\left(E_{A'}-E_{A}\right)\langle A'|U^{(2)}|A\rangle .
\end{eqnarray}
The general proof of Eqs. (\ref{30}), (\ref{31}) one can find in \cite{AndrLab}.

Insertion of the expressions for the electron (Eq. (\ref{19})) and photon (Eq. (\ref{29})) propagator into 
Eq. (\ref{28}) and integrating over time and frequency variables yields
\begin{eqnarray}
\label{32}
\Delta E^{(2)}_A=\frac{e^2}{2\pi}
\sum\limits_n\left(\frac{1-\vec{\alpha}_1\vec{\alpha}_2}{r_{12}}I_{nA}(r_{12})\right)_{AnnA}\, ,
\end{eqnarray}
\begin{eqnarray}
\label{33}
I_{nA}(r_{12}) = \int\limits_{-\infty}^{\infty}\frac{e^{i|\omega_1|r_{12}}d\omega}{E_n(1-i0)-E_A+\omega}\, ,
\end{eqnarray}
where $\vec{\alpha}$ (i=1,2) are the Dirac matrices acting on the different electron variables.

The exact evaluation of the integral (\ref{33}) in the complex $\omega$ plane  results in \cite{LabKlim}:
\begin{eqnarray}
\label{34}
I_{nA}(r_{12}) = \pi i \left(1+\frac{E_n}{|E_n|}\right)\left(1-\frac{\beta_{nA}}{|\beta_{nA}|}\right)e^{i|\beta_{nA}|r_{12}}+2i\frac{\beta_{nA}}{|\beta_{nA}|}\,
\left[ci\left(|\beta_{nA}|r_{12}\right)\,\sin\left(|\beta_{nA}|r_{12}\right)-si\left(|\beta_{nA}|r_{12}\right)\,\cos\left(|\beta_{nA}|r_{12}\right)\right]\, ,\nonumber \\
\end{eqnarray}
where $\beta_{nA}=E_n-E_A$ and the notations $si(x)$ and $ci(x)$ for the integral sine and cosine functions are employed. Then, according to Eq. (\ref{27}) the pure one-photon width $\Gamma^{(1)}_A$ can be 
represented in a closed form \cite{LabKlim}
\begin{eqnarray}
\label{35}
\Gamma^{(1)}_A=-\frac{e^2}{2}\sum\limits_n\left(1-\frac{\beta_{nA}}{|\beta_{nA}|}\right)\left(1+\frac{E_n}{|E_n|}\right)\left(\frac{1-\vec{\alpha}_1\vec{\alpha}_2}{r_{12}}\sin\left(|\beta_{nA}|r_{12}\right)\right)_{AnnA}\, ,
\end{eqnarray}
where the summation over $\vec{e}$ and integration over $\vec{\nu}$, remaining in Eq. (\ref{11}), is now performed. Note, that the summation in Eq. (\ref{35}) extends only over the electron states with energy $-m< E_n<E_A$.

The variable $\beta_{nA}r_{12}$ is of the order $\alpha Z$, so that in the nonrelativistic limit $\beta_{nA}r_{12}\ll 1$. Then expanding $\sin\left(|\beta_{nA}|r_{12}\right)$ in Eq. (\ref{35}) we can again make contact with
the nonrelativistic expression Eq. (\ref{14}), or Eq. (\ref{17}) for $\Gamma^{(1)}_A$.

\section{Decay width via the imaginary part of the Lamb shift: application of the adiabatic theory}

In this Section we will apply the Gell-Mann and Low adiabatic formula \cite{Gell} for the energy shift $\Delta E_A$ (Lamb shift)
of an excited atomic state $A$ due to the interaction with the vacuum of the radiation fields 
\begin{eqnarray}
\label{36}
\Delta E_A = \lim_{\eta\rightarrow 0}\frac{1}{2}i\eta\frac{e\frac{\partial}{\partial e}\langle A|\hat{S}_{\eta}|A\rangle }{\langle A|\hat{S}_{\eta}|A\rangle }\, .
\end{eqnarray}
for the evaluation of the imaginary part of the Lamb shift. 
The adiabatic $S$-matrix $\hat{S}_{\eta}$ differs from the usual $S$-matrix by the presence of the adiabatic (exponential) factor $e^{-\eta|t|}$ in each (interaction) vertex. It refers to the concept of adiabatic switching on and off the interaction introduced formally by the replacement
$\hat{H}_{{\rm int}}(t) \longrightarrow \hat{H}^\eta_{{\rm int}}(t) = e^{-\eta|t|}\,\hat{H}_{{\rm int}}(t)$. 
The symmetrized version of the adiabatic formula containing $S_{\eta}(\infty,-\infty)$, which is more convenient for the QED calculations, was proposed by Sucher \cite{Sucher}. The first application of the formula (\ref{36}) to calculations within bound-state QED was made in \cite{Lab}. In \cite{Lab} it was shown how to deal with the adiabatic exponential factor when evaluating the real part of corrections to the energy levels Eq. (\ref{36}) (see also \cite{LabKlim}). In this paper we will employ the same methods for evaluating the imaginary part of Eq. (\ref{36}).

For a free atom (or ion) in the state $|A\rangle$ interacting with the photon vacuum $|0_\gamma\rangle$ 
(i.e. $|A, 0_\gamma\rangle =|A\rangle|0_\gamma\rangle$
in the absence of external fields)  the complex energy correction 
Eq. (\ref{26}) contains only diagonal $S$-matrix elements of even order, since 
$\langle 0_\gamma|\hat{S}^{(1)}_{\eta}|0_\gamma\rangle = 
\langle 0_\gamma|\hat{S}^{(3)}_{\eta}|0_\gamma\rangle = 0$ etc. 
For the separation of the imaginary part of the energy shift $\Delta E_A^{(2i)}$ of a given order $2i$,  
it is more convenient to represent Eq. (\ref{36}) in terms of a perturbation series of the form
(up to terms $e^4$)  \cite{LabKlim} 
\begin{eqnarray}
\label{37}
\Delta E_A = \lim_{\eta\rightarrow 0} i\eta\, \left[\langle A|\hat{S}^{(2)}_{\eta}|A\rangle + 
\left(2 \langle A|\hat{S}^{(4)}_{\eta}|A\rangle - \langle A|\hat{S}^{(2)}_{\eta}|A\rangle^2\right)
\dots \right]\, .
\end{eqnarray}
For the adiabatic $\hat{S}_{\eta}$ matrix we used the standard expansion in powers of the interaction
constant $e$ 
\begin{eqnarray}
\label{38}
\hat{S}_{\eta}(\infty,-\infty)=1+\sum\limits_{i=1}^{\infty}\hat{S}^{(i)}_{\eta}(\infty,-\infty)
\end{eqnarray}
and can separate real and imaginary parts of the matrix elements at any given order of perturbation theory
\begin{eqnarray}
\label{39}
\langle A|\hat{S}^{(i)}_{\eta}|A\rangle  = Re\langle A|\hat{S}^{(i)}_{\eta}|A\rangle +iIm\langle A|\hat{S}^{(i)}_{\eta}|A\rangle\, .
\end{eqnarray}

The only one second-order term describes the pure one-photon decay width
\begin{eqnarray}
\label{40}
Im\Delta E_A^{(2)}=\lim_{\eta\rightarrow 0}\eta Re\langle A|\hat{S}_{\eta}^{(2)}|A\rangle\, .
\end{eqnarray}

Arranging all the terms of fourth order, which describe the pure two-photon decay width including 
- as we will see below - a part of the (one-loop) radiative corrections to the one-photon width, 
one obtains 
\begin{eqnarray}
\label{41}
Im\Delta E_A^{(4)} = \lim_{\eta\rightarrow 0} \eta \left[2 Re\langle A|\hat{S}_{\eta}^{(4)}|A\rangle +\left|\langle A|\hat{S}_{\eta}^{(2)}|A\rangle \right|^2-2\left(Re\langle A|\hat{S}_{\eta}^{(2)}|A\rangle \right)^2\right]\, ,
\end{eqnarray}
where the last two terms result from the expression $\langle A|\hat{S}_{\eta}^{(2)}|A\rangle^2$.

The total width $\Gamma_A$ of an excited electron state $A$ (specifying the initial state as 
$|A, 0_\gamma\rangle \equiv |A\rangle$) should follow (by definition) from the imaginary part 
of the total energy-shift via  
\begin{eqnarray}
\label{41bb}
\Gamma_A = -2 Im\Delta E_A\, ,
\end{eqnarray}
respectively, after perturbation expansion of $\Delta E_A$ (up to order $e^4$) as
\begin{eqnarray}
\label{41b}
\Gamma_A = \lim_{\eta\rightarrow 0} -2\eta \left[
Re \langle A|\hat{S}_{\eta}^{(2)}|A\rangle  + 
2 Re \langle A|\hat{S}_{\eta}^{(4)}|A\rangle 
+ \left|\langle A|\hat{S}_{\eta}^{(2)}|A\rangle \right|^2 - 
2\left(Re\langle A|\hat{S}_{\eta}^{(2)}|A\rangle \right)^2\right]\, .
\end{eqnarray}

The formulas (\ref{40}), (\ref{41}) and (\ref{41b}) will be employed in the next Sections for evaluating the pure one-photon and two-photon widths $\Gamma_A^{(1)}$ and $\Gamma_A^{(2)}$, respectively.

\section{Application of the "optical theorem"}

\subsection{Formulation of the "optical theorem" for the $S$-matrix elements}

The "optical theorem" is a consequence of the unitarity of the $S$-matrix and in the most general case 
can be formulated as follows (see, for example, \cite{bas59}).
First, we introduce the $\hat{T}$-matrix via the  definition
\begin{eqnarray}
\label{42}
\hat{S}=1+i\hat{T}\, .
\end{eqnarray}
For the diagonal matrix element of  Eq. (\ref{42}) it follows 
\begin{eqnarray}
\label{43}
\langle I|\hat{S}|I\rangle = 1+i\langle I|\hat{T}|I\rangle 
\end{eqnarray}
and 
\begin{eqnarray}
\label{43b}
Re \langle I|(1-\hat{S})|I\rangle = Im \langle I|\hat{T}|I\rangle \, .
\end{eqnarray}
Here $|I\rangle$ denotes the initial state of the (decaying) system ''atom + radiation field''. 
The wave function $|I\rangle $ is assumed to be normalized; in our later case of interest it refers to 
the wave function for the excited atom state and the photon vacuum. 
As a consequence of the unitarity relation $\hat{S}^\dagger\hat{S} = \hat{S}\hat{S}^\dagger = 1$ for the 
$S$-matrix, the "optical theorem" states
\begin{eqnarray}
\label{44b}
i \left(\hat{T} - \hat{T}^\dagger \right) = - \hat{T}^\dagger \hat{T} = - \hat{T}\hat{T}^\dagger\, ,
\end{eqnarray}
respectively, for the matrix elements
\begin{eqnarray}
\label{44}
2Im\langle I|\hat{T}|I\rangle =\sum\limits_F \left|\langle F|\hat{T}|I\rangle \right|^2\, .
\end{eqnarray}
The summation in Eq. (\ref{44}) runs over the complete set of final states $F$ allowed by the energy conservation law. Formally, the state $F=I$ is also included in this summation. The latter circumstance will be important for our further derivations. 
The initial and final states $|I\rangle$ and $|F\rangle$ are by definition eigenstates of the electron and photon number operator. 
We also mention at this point, that the ''optical theorem'' strictly holds for arbitrary not explicitly
time-dependent problems. Specific modifications are required in the presence of time-dependent
external field, which allow for electron-positron pair creation out of the Dirac vacuum. 
Under such conditions the $S$-matrix will become nonunitary \cite{fgs91}.

Performing perturbation expansion of the $T$-(respectively for the $S$-) matrix one derives immediately
from (\ref{44b}) and  (\ref{44}) the ''optical theorem'' for the $T$-matrix elements up to any (even) 
order $2i$ of perturbation theory 
\begin{eqnarray}
\label{44c}
2 Im \langle I|\hat{T}^{(2i)}|I\rangle &=& 
\sum\limits_F \left|\langle F|\hat{T}^{(i)}|I\rangle \right|^2 
 + \sum\limits_{F}\, \sum\limits_{j < i} 
2 Re\langle I|\hat{T}^{(j)\,\dagger}|F\rangle\langle F|\hat{T}^{(2i-j)}|I\rangle \, .
\end{eqnarray}
Depending on the physical process (respectively, the scenario) under consideration
one has to fix the number of electrons (atomic state) and photons (radiation field at zero temperature) 
in the inital and final state. However, the quantum numbers of electrons ($A$) and photons 
($\vec k, \vec e$) will vary of course. 
In case of the one-photon decay the summation over $F$ includes the summation over the final atomic states $A'$ as well as over the quantum numbers $\vec k, \vec e$ 
of the emitted photon. In case of the two-photon decay the summation over $F$ includes, apart from the summation over $A'$, also the summation (integration) over the characteristic quantum numbers  of the two emitted photons.

Expanding both sides of Eq. (\ref{43b}) into a perturbation series (with respect to $e$), 
we find for arbitrary orders $i=1, 2, \dots$ the relation 
\begin{eqnarray}
\label{45}
Re\langle I|\hat{S}^{(i)}|I\rangle =-Im\langle I|\hat{T}^{(i)}|I\rangle \, .
\end{eqnarray}
and thus the $S$-matrix form of the ''optical theorem'' corresponding to Eq. (\ref{44c})
\begin{eqnarray}
\label{44d}
- 2 Re \langle I|\hat{S}^{(2i)}|I\rangle &=& 
\sum\limits_F \left|\langle F|\hat{S}^{(i)}|I\rangle \right|^2 
 + \sum\limits_{F}\, \sum\limits_{j < i} 
2 Re \langle I|\hat{S}^{(j)\,\dagger}|F\rangle\langle F|\hat{S}^{(2i-j)}|I\rangle \, .
\end{eqnarray}
Then, collecting the second-order terms in Eqs. (\ref{44}) and (\ref{44d}) yields
\begin{eqnarray}
\label{46}
-2Re\langle I|\hat{S}^{(2)}|I\rangle =\sum\limits_{F\neq I}\left|\langle F|\hat{S}^{(1)}|I\rangle \right|^2.
\end{eqnarray}
Again, only nondiagonal matrix elements, like 
$\langle F|\hat{S}^{(1)}|I\rangle $ contribute. 

Collecting now the fourth-order terms, we find
\begin{eqnarray}
\label{47}
-2Re\langle I|\hat{S}^{(4)}|I\rangle =\left|\langle I|\hat{S}^{(2)}|I\rangle \right|^2+\sum\limits_{F\neq I}\left|\langle F|\hat{S}^{(2)}|I\rangle \right|^2+ \sum\limits_{F\neq I}
2Re\langle I|\hat{S}^{(1)}|F\rangle
\langle F|\hat{S}^{(3)}|I\rangle .
\end{eqnarray}
The last term in Eq. (\ref{47}) represents, evidently, the radiative corrections to the one-photon width. These corrections were evaluated by Barbieri and Sucher via direct evaluation of the corresponding imaginary part of the two-loop Lamb shift \cite{bas78}. Here we will not repeat these calculations within our approach. Note, that the term $F=I$ in the sum over $F$ is absent in this contribution, since $\langle I|\hat{S}^{(1)}|I\rangle =\langle I|\hat{S}^{(3)}|I\rangle =0$. 

\subsection{Application of the "optical theorem" to the adiabatic $S$-matrix elements}
As indicated above the adiabatic $S$-matrix $\hat{S}_\eta$ arises after introduction of the adiabatic
switching function $f(\eta) = e^{-\eta|t|}$ in the QED interaction Hamiltonian. 
Assuming that no dynamic excitations of the system takes place during switching on and off the interaction,
the adiabatic $S$-matrix remains unitary \cite{fgs91}, \cite{Berest}. 
Moreover, all observables calculated on the basis of the adiabatic approach should not  
depend on the specific form used for the adiabatic factor after the limiting process 
$\eta\rightarrow 0$ has been performed. 
Therefore, we will apply the "optical theorem" relations (\ref{46}) and (\ref{47}) to the adiabatic formulas (\ref{40}), (\ref{41}) and (\ref{41b}). In what follows, it will be necessary to fix not only the state of an electron in an atom, but also the number of the photons. 

Then from Eq. (\ref{40}), (\ref{41b}) and (\ref{47}) it follows for $I=A,0_\gamma$ (excited state, no photons)
for the pure one-photon width 
\begin{eqnarray}
\label{48}
\Gamma_A^{(1)}= 
\lim_{\eta\rightarrow 0} \eta\,\sum\limits_{F\neq A,0_\gamma}\left|\langle F|\hat{S}^{(1)}_{\eta}|A,0_\gamma\rangle\right|^2 
\end{eqnarray}
and for the  two-photon width 
\begin{eqnarray}
\label{49}
\Gamma_A^{(2)}= 
\lim_{\eta\rightarrow 0}\eta \left\lbrace 2\sum\limits_{F\neq A,0_\gamma}\left|\langle F|\hat{S}^{(2)}_{\eta}|A, 0_\gamma\rangle\right|^2+4\left(Re\langle A, 0_\gamma|\hat{S}^{(2)}_{\eta}|A, 0_\gamma\rangle\right)^2\right\rbrace \, .
\end{eqnarray}
The remaining term up to order $e^4$ containing radiative-correction effects 
\begin{eqnarray}
\Gamma_A^{{\rm rad}} = \lim_{\eta\rightarrow 0}\eta \,\sum\limits_{F\neq A,0_\gamma}\,
2 Re \langle A, 0_\gamma|\hat{S}^{(1)}|F\rangle\langle F|\hat{S}^{(3)}|A, 0_\gamma\rangle
\end{eqnarray}
will not be considered further, since we are aiming at the two-photon decay.

Employing now Eq. (\ref{46}), we can rewrite Eq. (\ref{49}) finally into the form
\begin{eqnarray}
\label{50}
\Gamma_A^{(2)}=\lim_{\eta\rightarrow 0} \eta \left\lbrace
2\sum\limits_{F\neq A,0_\gamma}\left|\langle F|\hat{S}^{(2)}_{\eta}|A, 0_\gamma\rangle\right|^2+
2\sum\limits_{2_\gamma}\left|\langle A, 2_\gamma|\hat{S}^{(2)}_{\eta}|A, 0_\gamma\rangle\right|^2+
\left(\sum\limits_{F'\neq A,0_\gamma}\left|\langle F'|\hat{S}^{(1)}_{\eta}
|A, 0_\gamma\rangle\right|^2\right)^2
 \right\rbrace .
\end{eqnarray}
In Eq. (\ref{50}) we have to distinguish between the final states ($F$) and ($F'$) 
for two-photon and for the one-photon transitions, respectively. 
It is important that the term $\left|\langle A, 0_\gamma|\hat{S}^{(2)}_{\eta}|A, 0_\gamma\rangle\right|^2$ has canceled out in Eq. (\ref{49}). The last but one term in (\ref{50}), corresponding to apparently nonphysical transition $A\rightarrow A+2\gamma$, but formally present in the sum over $F$ states, will indeed cancel out in the final expression (see Section 9). The notation $\sum\limits_{2_\gamma}$ means here the integration over the frequencies of two photons. 
In the next Section we will evaluate the one- and two-photon decay widths using Eqs (\ref{48}) and (\ref{50}).

\section{One-Photon decay width via the "optical theorem"}

We start with the evaluation of the decay width $\Gamma^{(1)}_A$ using Eq. (\ref{48}). First, we evaluate the matrix element $\langle A',\vec{k},\vec{e}|\hat{S}^{(1)}_{\eta}|A, 0_\gamma\rangle$ for the emission of the photon with momentum $\vec{k}$ and polarization $\vec{e}$.  This matrix element for the "normal" $S$-matrix was evaluated in Section 2. The corresponding adiabatic $S_{\eta}$-matrix element reads
\begin{eqnarray}
\label{51}
\langle A', \vec{k},\vec{e}|\hat{S}^{(1)}_{\eta}|A, 0_\gamma\rangle=e\int d^4x\bar{\psi}_{A'}(x)\gamma_{\mu}A^{*}_{\mu}(x)\psi_A(x)e^{-\eta |t|}.
\end{eqnarray}
Now the integration over the time variable yields essentially a representation of the $\delta$-function 
\begin{eqnarray}
\label{52}
\int\limits_{-\infty}^{\infty}dte^{i(E_A-E_{A'}-\omega)t-\eta |t|}=\frac{2\eta}{(\omega_{AA'}-\omega)^2+\eta^2} \equiv 2\pi\,\delta_\eta (\omega_{AA'}-\omega),
\end{eqnarray}
where $\lim\limits_{\eta\rightarrow 0}\delta_{\eta}(x)=\delta (x)$.
As the next step we perform the integration over the photon frequency. Taking Eq. (\ref{52}) by squre modulus, multiplying by $\omega$ and integrating, we obtain
\begin{eqnarray}
\label{53}
4\eta^2\int\limits_0^{\infty}\frac{\omega d\omega}{\left[(\omega_{AA'}-\omega)^2+\eta^2\right]^2}=4\eta^2\left\lbrace 
\frac{\pi\omega_{AA'}}{4\eta^3}+\frac{1}{2\eta^2}+\frac{\omega_{AA'}}{2\eta^3}arctg\left(\frac{\omega_{AA'}}{\eta}\right)
\right\rbrace 
\end{eqnarray}
Having in mind the limit $\eta\rightarrow 0$ we can replace Eq. (\ref{53}) by
\begin{eqnarray}
\label{54}
4\eta^2\int\limits_0^{\infty}\frac{\omega d\omega}{\left[(\omega_{AA'}-\omega)^2+\eta^2\right]^2}=\frac{2\pi\omega_{AA'}}{\eta}
\end{eqnarray}

It remains to multiply the result by the factor $(2\pi)^{-3}$ (see Eq. (\ref{9})), by the factor $(\sqrt{2\pi})$ (see Eq. (\ref{2})) and by the factor $\eta$ from Eq. (\ref{48}). The matrix element will be the same as in Eq. (\ref{4}) and we again arrive at Eq. (\ref{11}) with the summation over the electron states, lower by energy than the state $A$:
\begin{eqnarray}
\label{55}
\Gamma_A^{(1)}=\frac{e^2}{2\pi}
\sum\limits_{A'\,(E_{A'}<E_A)} \omega_{AA'} \sum\limits_{\vec{e}}\int d\vec{\nu} 
\left|\left( ( \vec{e}^{\,*}\vec{\alpha})e^{-i\vec{k}\vec{r}} \right)_{A'A} \right|^2
\end{eqnarray}
In the derivation above the manipulations with $\delta$-functions, like in Section 2, 
have been avoided. 
Multiplying the result by the adiabatic parameter $\eta$ in Eq. (\ref{48}) plays the same role as dividing the result by the time $T$ in Section 2: the adiabatic factor $\eta$ has the dimensionality
$s^{-1}$. Note, that in this approach the automatic exclusion (like in Eq. (\ref{35})) of the transitions to the states higher than $A$ in the summation over $F$ in Eq. (\ref{48}) does not occur and we have to refer to the energy conservation law to avoid them.

\section{Two-photon decay width via "optical theorem" in the absence of cascades}

\subsection{Evaluation of the two-photon decay width}

In this Section we will evaluate the two-photon decay width $\Gamma^{(2)}_A$ using Eq. (\ref{50}). We start with the first term in the curly brackets in Eq. (\ref{50}). The $S$-matrix elements corresponding to the emission of the two photons $\vec{k}$, $\vec{e}$ and $\vec{k}'$, $ \vec{e}\,'$ look like
\begin{eqnarray}
\label{56}
\langle A', \vec{k}', \vec{e}\,';\vec{k},\vec{e}|\hat{S}^{(2)}_{\eta}|A, 0_\gamma\rangle_a=e^2\int d^4x_1d^4x_2\left(\bar{\psi}_{A'}(x_1)\gamma_{\mu_1}A^{\vec{k}', \vec{e}\,'\,\,*}_{\mu_1}(x_1)e^{-\eta |t_1|}S(x_1x_2)\gamma_{\mu_2}A^{\vec{k},\vec{e}\,\,*}_{\mu_2}(x_2)e^{-\eta |t_2|}\psi_A(x_2)\right),
\end{eqnarray}
\begin{eqnarray}
\label{57}
\langle A', \vec{k},\vec{e};\vec{k}', \vec{e}\,'|\hat{S}^{(2)}_{\eta}|A, 0_\gamma\rangle_b=e^2\int d^4x_1d^4x_2\left(\bar{\psi}_{A'}(x_1)\gamma_{\mu_1}A^{\vec{k},\vec{e}\,\,*}_{\mu_1}(x_1)e^{-\eta |t_1|}S(x_1x_2)\gamma_{\mu_2}A^{\vec{k}', \vec{e}\,'\,\,*}_{\mu_2}(x_2)e^{-\eta |t_2|}\psi_A(x_2)\right),
\end{eqnarray}
where the electron propagator $S(x_1x_2)$ is defined by Eq. (\ref{19}) and the indices $a$, $b$ correspond to the Feynman graphs Figs 2a, 2b, respectively. The integration over $t_2$ in Eq. (\ref{56}) results
\begin{eqnarray}
\label{58}
\int\limits_{-\infty}^{\infty}dt_2e^{-i\left(\omega_1+E_A-\omega\right)t_2-\eta |t_2|}=\frac{2\eta}{(\omega_1+E_A-\omega)^2+\eta^2}
\end{eqnarray}
and similarly looks the integration over $t_1$ in Eq. (\ref{56})
\begin{eqnarray}
\label{59}
\int\limits_{-\infty}^{\infty}dt_2e^{i\left(\omega_1+E_{A'}+\omega'\right)t_1-\eta |t_1|}=\frac{2\eta}{(\omega_1+E_{A'}+\omega')^2+\eta^2}
\end{eqnarray}
The next step is the integration over $\omega_1$ in Eq. (\ref{56}) which has to be performed in the complex $\omega_1$-plane. We have to evaluate the integral (considering the $i0$-perscription in the energy 
denominator of the electron propagator)
\begin{eqnarray}
\label{60}
I_{\eta}\equiv 4\eta^2\int\limits_{-\infty}^{\infty}\frac{d\omega_1}{\left[(\omega_1+E_{A}-\omega)^2+\eta^2\right]\left[(\omega_1+E_{A'}+\omega')^2+\eta^2\right] \left[E_n(1-i0)+\omega_1\right]}.
\end{eqnarray}
In what follows in this Section we will restrict ourselves to the nonrelativistic limit of Eqs (\ref{56}), (\ref{57}), since we are most interested in the hydrogen case. 
Then we can fully neglect the sum over the negative-energy states in the electron propagator (\ref{19}). 
Note, that it is possible while we are using "velocity" form for the matrix elements of the photon emission operator. In the "length" form it would not be the case \cite{Akhiezer}. 
So we can close the integration contour in the lower half-plane, where only two poles are located: $\omega_1^{(1)}=-E_A+\omega+i\eta $ and $\omega_1^{(2)}=-E_{A'}-\omega'+i\eta $. In the absence of cascades, the energy denominators $(E_n-E_A+\omega-i\eta )^{-1}$ and $(E_n-E_{A'}+\omega'-i\eta )^{-1}$ are nonsingular and we can omit the imaginary parts $i\eta$ in these denominators. Moreover, using energy conservation law condition, $E_{A'}+\omega'=E_A-\omega$ we can consider both denominators as being equal. 
Then, collecting the factors ($1/2\pi$ from Eq. (\ref{19}), $-2\pi$ from the Cauchy formula for the contour integration in the clockwise direction) yields for the amplitude Eq. (\ref{56})
\begin{eqnarray}
\label{61}
\langle A', \vec{k}', \vec{e}\,';\vec{k},\vec{e}|\hat{S}^{(2)}_{\eta}|A, 0_\gamma\rangle_a=-\frac{4\eta}{\left[(\omega_0-\omega-\omega')^2+4\eta^2\right]}\sum\limits_n\frac{\langle A'|\vec{\alpha}\vec{A}_{\vec{k}', \vec{e}\,'}^{*}|n \rangle\langle n| \vec{\alpha}\vec{A}_{\vec{k},\vec{e}}^{*}|A\rangle }{E_n-E_A+\omega}
\end{eqnarray}
Adding the similar contribution from Eq. (\ref{57}) leads to
\begin{eqnarray}
\label{62}
\langle A'|\hat{S}^{(2)}_{\eta}|A, 0_\gamma\rangle_{a+b}=\hspace{5cm}
\\
\nonumber
=-\frac{4\eta}{\left[(\omega_0-\omega-\omega')^2+4\eta^2\right]}
\left[\sum\limits_n
\frac{\langle A'|\vec{\alpha}\vec{A}_{\vec{k}', \vec{e}\,'}^{*}|n \rangle\langle n| \vec{\alpha}\vec{A}_{\vec{k},\vec{e}}^{*}|A\rangle }{E_n-E_A+\omega}+\sum\limits_n
\frac{\langle A'|\vec{\alpha}\vec{A}_{\vec{k},\vec{e}}^{*}|n \rangle\langle n| \vec{\alpha}\vec{A}_{\vec{k}', \vec{e}\,'}^{*}|A\rangle }{E_n-E_{A'}-\omega}\right]\, .
\end{eqnarray}
In the matrix elements of the photon emission operators, unlike the $S$-matrix elements, we retain the 
shorthand notation $|A\rangle $ instead of $|A, 0_\gamma\rangle$, since in this case it cannot lead to any
misunderstandings.

Insertion of Eq. (\ref{62}) into the first term on the right-hand side of Eq. (\ref{50}) and summation (integration) over quantum numbers of the final-state particles yields in the nonrelativistic limit 
\begin{eqnarray}
\label{63}
\Gamma^{(2)}_{A}(1st\,\, term)=\lim_{\eta\rightarrow 0}\, 2\eta\, (\sqrt{2\pi})^4\frac{e^4}{(2\pi)^6}\sum\limits_{\vec{e}}\sum\limits_{ \vec{e}\,'}\int d\vec{\nu}d\vec{\nu}'\int\omega d\omega\int\omega'd\omega'\times
\\
\nonumber
\times\frac{(4\eta)^2}{\left[(\omega_0-\omega-\omega')^2+4\eta^2\right]^2}\left|\sum\limits_n
\frac{\langle A'|\vec{\alpha}\vec{A}_{\vec{k}', \vec{e}\,'}^{*}|n \rangle\langle n| \vec{\alpha}\vec{A}_{\vec{k},\vec{e}}^{*}|A\rangle }{E_n-E_A+\omega}+\sum\limits_n
\frac{\langle A'|\vec{\alpha}\vec{A}_{\vec{k},\vec{e}}^{*}|n \rangle\langle n| \vec{\alpha}\vec{A}_{\vec{k}', \vec{e}\,'}^{*}|A\rangle }{E_n-E_{A'}-\omega}\right|^2\, .
\end{eqnarray}
The integrations over $\omega$, $\omega'$ in Eq. (\ref{62}) can be performed using exactly 
the same standard QED procedure as in Section III. We first integrated over $\omega'$, using the
$\delta$-function in Eq. (\ref{20}), i.e. over the interval $(\infty, 0)$, or even over $(\infty,-\infty)$ what is actually equivalent in this case. The second integration over $\omega$ was performed over the interval $(0,\omega_0)$ (see Eq. (\ref{26})).

Let us adopt here this procedure within adiababtic $S_{\eta}$-matrix approach. 
Integrating Eq. (\ref{63}) over $\omega'$ according to Eqs (\ref{53}) and (\ref{54}) leads to 
\begin{eqnarray}
\label{64}
(4\eta)^2\int\limits_0^{\infty}\frac{\omega'd\omega'}{\left[(\omega_0-\omega-\omega')^2+4\eta^2\right]^2}=\frac{\pi(\omega_0-\omega)}{\eta}.
\end{eqnarray}
Then the integration over the emitted photon directions and summation over the polarizations yields
in the nonrelativistic limit (in the "velocity" form):
\begin{eqnarray}
\label{65}
\Gamma^{(2)}_{A}(1st\,\, term)=\frac{4e^4}{9\pi}\int\limits_0^{\omega_0}\omega (\omega_0-\omega)d\omega \sum\limits_{i,k=1}^3\left|\left(U_{ik}(\omega)\right)_{A'A}\right|^2,
\end{eqnarray}
\begin{eqnarray}
\label{66}
\left(U_{ik}\right)_{A'A}=\sum\limits_n
\frac{\langle A'|p_i|n \rangle\langle n| p_k|A\rangle }{E_n-E_A+\omega}+\sum\limits_n
\frac{\langle A'|p_k|n \rangle\langle n| p_i|A\rangle }{E_n-E_{A'}-\omega},
\end{eqnarray}
where $p_i\equiv(\vec{p})_i$. For $A=2s$, $A'=1s$ Eqs. (\ref{65}), (\ref{66}) go over to Eqs (\ref{23})-(\ref{25}), if we use again the quantum mechanical relation (\ref{15}).

\subsection{Cancellation of singularities}

Apart from the first term in Eq. (\ref{50}) there are two additional terms which contain the singularities with respect to the adiabatic parameter $\eta$ in the limit $\eta\rightarrow 0$. In this Subsection we will show that these singularities exactly cancel. We start with the last term in the right-hand side of Eq. (\ref{57}). In the absence of cascades, i.e. when there are no energy levels between the initial state $A$ and the final state $A'$, the only term in the sum over $F'$ is $F'=A',1_\gamma$. Then, after summation over the emitted photon polarization and integration over th emitted photon directions, repeating the derivations performed in Section 8, we obtain the result
\begin{eqnarray}
\label{67}
\lim_{\eta\rightarrow 0}\,\eta\,\left(\sum\limits_{F'\neq A, 0}\left|\langle F'|\hat{S}^{(1)}|A, 0_\gamma\rangle\right|^2\right)^2=\lim_{\eta\rightarrow 0}\frac{1}{\eta}\left(\Gamma_A^{(1)}\right)^2.
\end{eqnarray}
This divergent term can be canceled only by the "unphysical" contribution $2\sum\limits_{2\gamma}\left|\langle A, 2_\gamma|\hat{S}^{(2)}_{\eta}|A, 0_\gamma\rangle\right|^2$ in the right-hand side of Eq. (\ref{50}). The latter one looks exactly like Eq. (\ref{63}) if $A'$ is
replaced by $A$. Setting $n=A'$ in the sum over $n$ in the expression 
$2\langle A, 2_\gamma|\hat{S}_{\eta}^{(2)}|A, 0_\gamma\rangle$ will give the same set of the matrix elements
as in Eq. (\ref{67}). This contribution also appears to be divergent like $\eta^{-1}$ in the limit $\eta\rightarrow 0$ and cancels the divergency Eq. (\ref{67}). 

To demonstrate this, we write down the expression
\begin{eqnarray}
\label{67a}
\lim_{\eta\rightarrow 0}\,2\eta\,\sum\limits_{2\gamma}\left|\langle A, 2_\gamma|\hat{S}^{(2)}_{\eta}|A, 0_\gamma\rangle\right|^2=\lim_{\eta\rightarrow\, 0}2\eta\,(2\pi)^2\frac{e^4}{(2\pi)^6}\sum\limits_{\vec{e}}\sum\limits_{ \vec{e}\,'}\int d\vec{\nu}d\vec{\nu}'\int\omega d\omega\int\omega'd\omega'\times
\nonumber
\\
\frac{(4\eta)^2}{\left[(\omega_0-\omega-\omega')^2+4\eta^2\right]^2}\left|
\frac{\langle A|\vec{\alpha}\vec{A}_{\vec{k}', \vec{e}\,'}^{*}|A' \rangle\langle A'|\vec{\alpha}\vec{A}_{\vec{k},\vec{e}}^{*}|A\rangle }{E_{A'}-E_A+\omega+i\eta}+
\frac{\langle A|\vec{\alpha}\vec{A}_{\vec{k},\vec{e}}^{*}|A' \rangle\langle A'|\vec{\alpha}\vec{A}_{\vec{k}', \vec{e}\,'}^{*}|A\rangle }{E_{A'}-E_{A}-\omega+i\eta}\right|^2\, .
\end{eqnarray}
In this case, unlike (\ref{62}), we now keep $i\eta$ in the energy denominators, in order to keep trace of 
all the divergences. 
The integration over $\omega$ is exactly performed like in Eq. (\ref{64}) with the result
\begin{eqnarray}
\label{67b}
\lim_{\eta\rightarrow 0}\, 2\eta\,\sum\limits_{2\gamma}\left|\langle A, 2_\gamma|\hat{S}^{(2)}_{\eta}|A, 0_\gamma\rangle\right|^2 =
\lim_{\eta\rightarrow 0}\, 2\eta\, \frac{e^4}{(2\pi)^4}\sum\limits_{\vec{e}, \vec{e}\,'}\int d\vec{\nu}d\vec{\nu}'4
\left|\langle A|\vec{\alpha}\vec{A}_{\vec{k},\vec{e}}^{*}|A' \rangle\right|^2\left|\langle A'|\vec{\alpha}\vec{A}_{\vec{k}', \vec{e}\,'}^{*}|A\rangle \right|^2\times
\nonumber
\\
\times\left(-\frac{\pi}{\eta}\right)\int\omega^2d\omega\left|\frac{1}{E_{A'}-E_A+\omega+i\eta}+\frac{1}{E_{A'}-E_A-\omega+i\eta}\right|^2\hspace{2cm}
\end{eqnarray}

After summation over the polarizations and integration over the emission angels and transforming the
expression in $|...|^2$, we get
\begin{eqnarray}
\label{67c}
\lim_{\eta\rightarrow 0}\, 2\eta\, \sum\limits_{2\gamma}\left|\langle A, 2_\gamma|\hat{S}^{(2)}_{\eta}|A, 0_\gamma\rangle\right|^2 =
-\lim_{\eta\rightarrow 0}\frac{2}{\eta}\left(\Gamma^{(1)}_A\right)^2F(\eta)\, ,
\end{eqnarray}
where the function 
\begin{eqnarray}\label{67d}
F(\eta)=\int\frac{\omega^2d\omega}{(\omega_0^2-\eta^2-\omega^2)^2+4\eta^2\omega^2_0}
\end{eqnarray}
remains to be calculated. 
In this case we have to evaluate the integral over $\omega$ in the same way as for deriving
the expression (\ref{67}), i.e. integrating over the frequency interval $(0,\infty)$. 
However, it will be more convenient to extend formally the integration over the entire 
interval $(\infty,-\infty)$ as it was done, for example, in \cite{Berest}. 
Then the integration can be performed in the complex plane:
\begin{eqnarray}
\label{67da}
F(\eta)&=&\int\limits_{-\infty}^{\infty}\frac{\omega^2d\omega}{(\omega_0^2-\eta^2-\omega^2+2i\eta\omega_0)(\omega_0^2-\eta^2-\omega^2-2i\eta\omega_0)}\nonumber \\
&=& \int\limits_{-\infty}^{\infty}\frac{\omega^2d\omega}{[(\omega_0 + i\eta + \omega)
(\omega_0 + i\eta - \omega)][(\omega_0 - i\eta + \omega)(\omega_0 - i\eta - \omega)]}\, .
\end{eqnarray}
The denominator in Eq. (\ref{67da}) contains 4 single poles: one pole in each quadrant of the complex plane. 
%
However, we have to recall that two of these poles originate from the negative (i.e. unphysical) $\omega$ values. Therefore we have to omit their contribution. Then the evaluation of the integral (\ref{67da}) yields with the aid of Cauchy's theorem 
\begin{eqnarray}\label{67g}
F(\eta)=\frac{2\pi i}{4 i \eta}=\frac{\pi}{2\eta}
\end{eqnarray}
Inserting this result into Eq. (\ref{67c}) we find that this result exactly cancels the divergent contribution Eq. (\ref{67}). What concerns the terms with $n\neq A'$ in the sum over $n$ in the expression for $2\langle A, 2_\gamma|\hat{S}^{(2)}_{\eta}|A, 0_\gamma\rangle$, we will avoid them referring to the energy conservation law, exactly like we did it in the Section 8 for the one-photon transition. Aiming to cancel the contributions from Eq. (\ref{67}) and from $2\langle A, 2_\gamma|\hat{S}^{(2)}_{\eta}|A, 0_\gamma\rangle$ we have to treat both contributions in the same way.

Thus, the second and the third terms in Eq. (\ref{50}) in the absence of cascades cancel each other and the two-photon width is given exclusively by the first term in Eq. (\ref{50}): $\Gamma^{(2)}_A(1st\,\, term)=\Gamma_A^{(2)}$. The expression (\ref{66}) for $\Gamma_A^{(2)}$ coincides exactly with the standard QED expression for the two-photon decay width in the absence of the cascades. In case of the $2s$-level in hydrogen we return to the expression (\ref{23}) for $n=2$.

\section{Two-photon decay width via "optical theorem" in the presence of cascades}


The derivation of the expression for the two-photon decay width in the presence of the cascades does not differ, in principle, from the derivation of Section IX. The expression (\ref{63}) holds in this case 
as well. We will assume that there exists only one cascade channel (for example, $3s-2p-1s$ in case of the decay of $3s$-level in hydrogen). The only difference is the existence of an additional resonance by $n=R$ in the sum over $n$ in Eq. (\ref{63}): $\omega=E_A-E_R$ ($R$ is the resonance state). Due to the energy conservation this implies also the existence of another resonance (lower branch of the cascade): $\omega'=E_R-E_{A'}$. Now Eq. (\ref{67}) contains two divergent (like $\eta^{-1}$ by $\eta\rightarrow 0$) terms: by $F'=A', 1_\gamma$ (this divergency is the same as in case of the absence of cascades) and by $F'=R,1_\gamma$ (this is the additionaö divergency connected with the existence of the cascade). 
The former divergency is compensated by the term $n=A'$ in the "unphysical contribution" $2\langle A2_{\gamma}|\hat{S}^{(2)}_{\eta}|A, 0_\gamma\rangle$ and the latter one is compensated in the same way by the term $n=R$ in the expression for $2\langle A2_\gamma|\hat{S}{(2)}_{\eta}|A, 0_\gamma\rangle$. 
However, in the presence of the cascade a third divergency arises directly in the summation over $n$ 
(for $n=R$) in the expression (\ref{67a}). This divergency cannot be canceled by any counterterm since it is proportional to a special product of matrix element: $\left|\langle A'|p_i|R><R|p_k|A\rangle \right|^2$. No counterterm in Eq. (\ref{50}) contains such a product. 
This remaining divergency can only be removed by taking into account 
(at given order of perturbation theory) the radiative corrections 
to the level width. Equivalently, one may introduce the level
width as it was done in Section IV within the framework of standard QED. Thus, evaluating the two-photon decay width in the presence of the cascades, we again return back to the same expressions and the same problems which were discussed in Section IV.

Our analysis of the evaluation of the two-photon decay width in the presence of cascades via the imaginary part of the Lamb shift, performed in Sections IX, X contradicts to Jentschura's "alternative" approach \cite{Jent1}-\cite{Jent3}. First, the expression (\ref{65}) contains the square modulus $\left|\left(U_{ik}(\omega)\right)_{A'A}\right|^2$, not the squared matrix element $\left(\left(U_{ik}(\omega)\right)_{A'A}\right)^2$ as used in Jentschura's calculation. 
Second, there is no chance to compensate the cascade divergency in Eq. (\ref{63}) within the evaluation of the imaginary part of the Lamb shift via the adiabatic $S$-matrix approach. 
In Jentschura's evaluation of the two-photon decay width in the presence of cascades the integral 
remains finite without introduction of the level width. 
We claim that the cascade contribution to the two photon decay width 
remains infinite without an introduction of the level widths in the energy denominators (via partial resummation of radiative corrections).

\section{Evaluation of the two-photon width for the $3s$ level in hydrogen}

As a consequense of the studies presented in Sections 5-10, we return now to the standard QED expression for the two-photon decay of $3s$-level in hydrogen (Eqs (\ref{23})-(\ref{25}) with $n=3$) and to the 
prescription for the employment of the formulas (\ref{23})-(\ref{25}) given in Section 4. Inserting Eq. (\ref{23}) into Eq. (\ref{26}) and retaining only the resonant term within the second and fourth frequency intervals, will yield the cascade contribution to the total two-photon decay rate of the $3s$-level. Taking the ratio to the total width of the $3s$-level $\Gamma_{3s}$ we wil obtain the absolute probability or branching ratio $W^{(cascade)}_{3s;1s}/\Gamma_{3s}\equiv b^{(cascade)}_{3s-2p-1s}$ for the cascade transition. The contributions to $b^{(cascade)}_{3s-2p-1s}$ from the intervals (I), (III), (V) are assumed to be zero. The cascade contribution of the $3s$-level results (in the "length" form)
\begin{eqnarray}
\label{68}
W^{({\rm cascade}\, 1\gamma)}_{3s;1s}=\frac{4}{27\pi}\int\limits_{({\bf  II})}\omega^3(\omega_0-\omega)^3\left|\frac{\langle R_{3s}(r)|r|R_{2p}(r)\rangle\langle R_{2p}(r')|r'|R_{1s}(r')\rangle}{E_{2p}-E_{3s}+\omega-\frac{i}{2}\Gamma}\right|^2d\omega + 
\\
\nonumber
+\frac{4}{27\pi}\int\limits_{({\bf  IV})}\omega^3(\omega_0-\omega)^3\left|\frac{\langle R_{3s}(r)|r|R_{2p}(r)\rangle\langle R_{2p}(r')|r'|R_{1s}(r')\rangle}{E_{2p}-E_{1s}-\omega-\frac{i}{2}\Gamma_{2p}}\right|^2d\omega .
\end{eqnarray}

According to the discussion in Section 4 the "pure" two-photon decay probabilities within each interval, defined in Section 4, look like
\begin{eqnarray}
\label{69}
dW_{3s;1s}^{(\rm{pure} 2\gamma)}&=&\frac{4}{27\pi}\omega^3(\omega_0-\omega)^3\left[S_{1s;3s}^{(2p)}(\omega)+S_{1s;3s}(\omega_0-\omega)\right]^2d\omega, \,\, \omega\in {\bf II}
\end{eqnarray}
\begin{eqnarray}
\label{70}
dW_{3s;1s}^{(\rm{pure} 2\gamma)}&=&\frac{4}{27\pi}\omega^3(\omega_0-\omega)^3\left[S_{1s;3s}(\omega)+S_{1s;3s}^{(2p)}(\omega_0-\omega)\right]^2d\omega, \,\, \omega\in {\bf IV}
\end{eqnarray}
\begin{eqnarray}
\label{71}
dW^{(\rm{pure} 2\gamma)}_{3s;1s} &=& \frac{4}{27\pi}\omega^3(\omega_0-\omega)^3\left[S_{1s;3s}(\omega)+S_{1s;3s}(\omega_0-\omega)\right]^2d\omega, \,\, \omega \in {\bf I, III, V} \, .
\end{eqnarray}
Here $S_{1s;3s}^{(2p)}(\omega)$ is the expression (\ref{2}) with the $n=2$ term 
being excluded.

Unlike cascade, all the intervals contribute to the "pure" two-photon transition. The branching ratio for this transition $3s\rightarrow 2\gamma +1s$ appears to be
\begin{eqnarray}
\label{72}
b^{(\rm{pure} 2\gamma)}_{3s-1s} = \frac{1}{2}\frac{1}{\Gamma_{3s}}\int\limits_0^{\omega_0}dW^{(pure 2\gamma)}_{3s;1s}(\omega)\, .
\end{eqnarray}
It remains to introduce the interference contribution. This contribution we consider only for the 2nd and 4th intervals. The corresponding frequency distribution functions are given by 
\begin{eqnarray}
\label{73}
dW^{(\rm{inter})1}_{3s;1s}=\frac{4\omega^3(\omega_0-\omega)^3}{27\pi}Re\left[\frac{\langle R_{3s}(r)|r|R_{2p}(2r)\rangle\langle R_{2p}(r')|r'|R_{1s}(r')\rangle}{E_{2p}-E_{3s}+\omega-\frac{i}{2}\Gamma}\right]\left[S_{1s;3s}^{(2p)}(\omega)+S_{1s;3s}(\omega_0-\omega)\right]d\omega
\end{eqnarray}
\begin{eqnarray}
\label{74}
dW^{(\rm{inter})2}_{3s;1s}=\frac{4\omega^3(\omega_0-\omega)^3}{27\pi}Re\left[\frac{\langle R_{3s}(r)|r|R_{2p}(2r)\rangle\langle R_{2p}(r')|r'|R_{1s}(r')\rangle}{E_{2p}-E_{1s}-\omega-\frac{i}{2}\Gamma}\right]\left[S_{1s;3s}(\omega)+S_{1s;3s}^{(2p)}(\omega_0-\omega)\right]d\omega
\end{eqnarray}
and branching ratio results as 
\begin{eqnarray}
\label{75}
b^{(\rm{inter})}_{3s;1s} = \frac{1}{2\Gamma_{3s}}\int\limits_{({\bf II})}dW^{(\rm{inter})1}_{3s;1s}+\frac{1}{2\Gamma_{3s}}\int\limits_{({\bf IV})}dW^{(\rm{inter})2}_{3s;1s}.
\end{eqnarray}

The results of our calculations are presented in Table 1. 
It is convenient to define the size $\Delta \omega$ of the second interval as 
multiples $l$ of the widths $\Gamma$, i.e.   
$\Delta \omega = 2l\Gamma$ and for the fourth interval 
as $\Delta \omega = 2l\Gamma_{2p}$, respectively. In Table 1 numbers are given for different values of $l$ ranging from 
$l\simeq 10^5$ up to $l\simeq 10^7$. The upper bound of interval {\bf II} equals  $\omega_1+l\Gamma=\frac{5}{72}+l\Gamma$ (in a.u.), while the lower bound of interval 
{\bf IV} equals $\omega_2-l\Gamma_{2p}=\frac{3}{8}-l\Gamma_{2p}$. 
The different lines of the Table 1 present branching ratios and transition rates 
of the "pure" two-photon and "interference" channel, respectively. For the more detailed analysis the contributions of the "pure" two-photon transition rate for the each frequency interval are also compiled. The branching ratio and the transition rate for the cascade contribution can be obtained from the relation  $b^{(\rm{cascade})}_{3s-2p-1s} + b^{(\rm{pure}2\gamma)}_{2s;1s} + b^{(\rm{inter})}_{3s;1s}=1$. This relation is sutisfied with high accuracy since the only decay channel neglected is the very weak direct 1-photon $M1$ transition $3s\rightarrow 1s+\gamma$. From the Table 1 we can draw the following conlusions: as in the case of the HCI \cite{LabShon}, the "pure" two-photon and cascade contributions to the total decay rate appear to be inseparable. Changing the interval size $\Delta\omega$, 
we obtain quite different values for $dW^{(\rm{pure} 2\gamma)}_{3s;1s}$ ranging
from $202.16\, s^{-1}$ (for $l = 10^4$) up to $7.9385\, s^{-1}$ 
(for $l = 1.00256\cdot 10^7$).

Moreover, in our calculations - depending on the size of the interval - the interference contribution also can become quite large, comparable in magnitude with the "pure" two-photon
contribution. Thus, we demonstrated that even the 
order of magnitude of the "pure" two-photon decay rate for the $3s$-state in hydrogen can not be predicted reliably.

Earlier the result $8.2196$ $s^{-1}$ for the "pure" two-photon decay of the  $3s$-level was reported in \cite{cea86} and confirmed in \cite{fsm88}. However, as it was pointed out in \cite{Chluba} in both papers \cite{cea86}, \cite{fsm88} the summation over the intermediate states was not performed properly. The "nonresonant" contribution $10.556$ $s^{-1}$ deduced in \cite{Chluba}, which plays the role of the "pure" two-photon decay rate is well within the range of our values given Table 1. However, the result $2.08$ $s^{-1}$ obtained for the "pure" two-photon decay rate in \cite{jas08} is in strong contradiction with the present analysis.

Very recently, a paper \cite{Amaro} did arrive where both the standard QED
approach, based on the line profile theory (\cite{Drake}-\cite{LabShon}) and the
"alternative" approach based on the two-loop Lamb shift theory (\cite{Jent1}-\cite{Jent3})
were applied to the calculation of the two-photon transition in hydrogen.
A reasonable agreement between the two methods was found. However, from
the derivations in our present paper it follows that the employment of the
Lamb shift imaginary part gives exactly the same results as the standard
QED approach. The difference between the "standard" and the "alternative"
methods is due to the use of the squared amplitude instead of square
modulus in Jentschura's calculation. To our mind this replacement is
unacceptable and cannot be justified within QED.

\section{Conclusion}
In this paper we developed a method for the calculation of the two-photon decay rates, based on the evaluation of the imaginary part of the Lamb shift with employment of the adiabatic $S$-matrix theory and 
the "optical theorem". We have  shown that the results of such calculations coincide exactly with the standard QED approach also in the presence of cascades.

We demonstrated that a strict separation of the "pure" two-photon and cascade contributions for $3s$-level decay in hydrogen is impossible. Moreover, we show that even the approximate separation of these two decay channels cannot be achieved with an accuracy, required in modern astrophysical investigations (i.e. at 1\% level) of the recombination history of hydrogen in the early Universe.

As a possible solution of the problem with respect to astrphysical needs, we would suggest to rewrite the
basic evolution equation for the number of the hydrogen atoms in a certain excited state (i.e. Eq. (2) in \cite{Hirata}) in a way which does not distinguish "pure" two-photon decays and cascades.

\begin{center}
Acknowledgments
\end{center}
The authors are grateful to R. A. Sunyaev and J. Chluba for stimulating interest in the problem and for many valuable discussions. The authors acknowledge financial support from DFG and GSI. 
The work was also supported by RFBR grant Nr. 08-02-00026. 
The work of D. S. was supported by the Non-profit Foundation “Dynasty” (Moscow). L. L. and D. S. acknowledge also the support by the Program of development of scientific potential of High School, Ministry of Education and Science of Russian Federation, grant $\aleph$2.1.1/1136.

\begin{figure}[ht]
\centerline{\mbox{\epsfxsize=60pt \epsffile{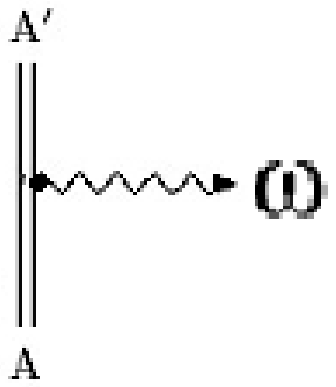}} }
\end{figure}
\small{Fig. 1. The Feynman graph corresponding to the photon emission in an one-electron atom. The double solid line describes the electron in the firld of nucleus (Furry picture), the wavy line with the arro at the end describes the emitted photon. The indices $A$ and $A'$ refer to the quantum numbers of the initial and final states of an atom, $\omega$ denotes the frequency of the emitted photon.
}

\begin{figure}[ht]
\centerline{\mbox{\epsfxsize=150pt \epsffile{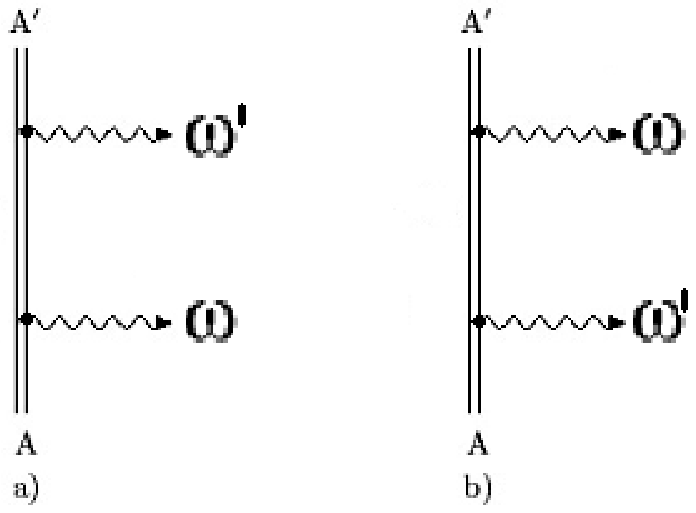}} }
\end{figure}
\small{Fig. 2. The Feynman graphs corresponding to the two-photon transition $A\rightarrow A'+2\gamma$. By $\omega$, $\omega'$ we denote the frequencies of the emitted photons.
}
\newpage
\begin{figure}[ht]
\centerline{\mbox{\epsfxsize=200pt \epsffile{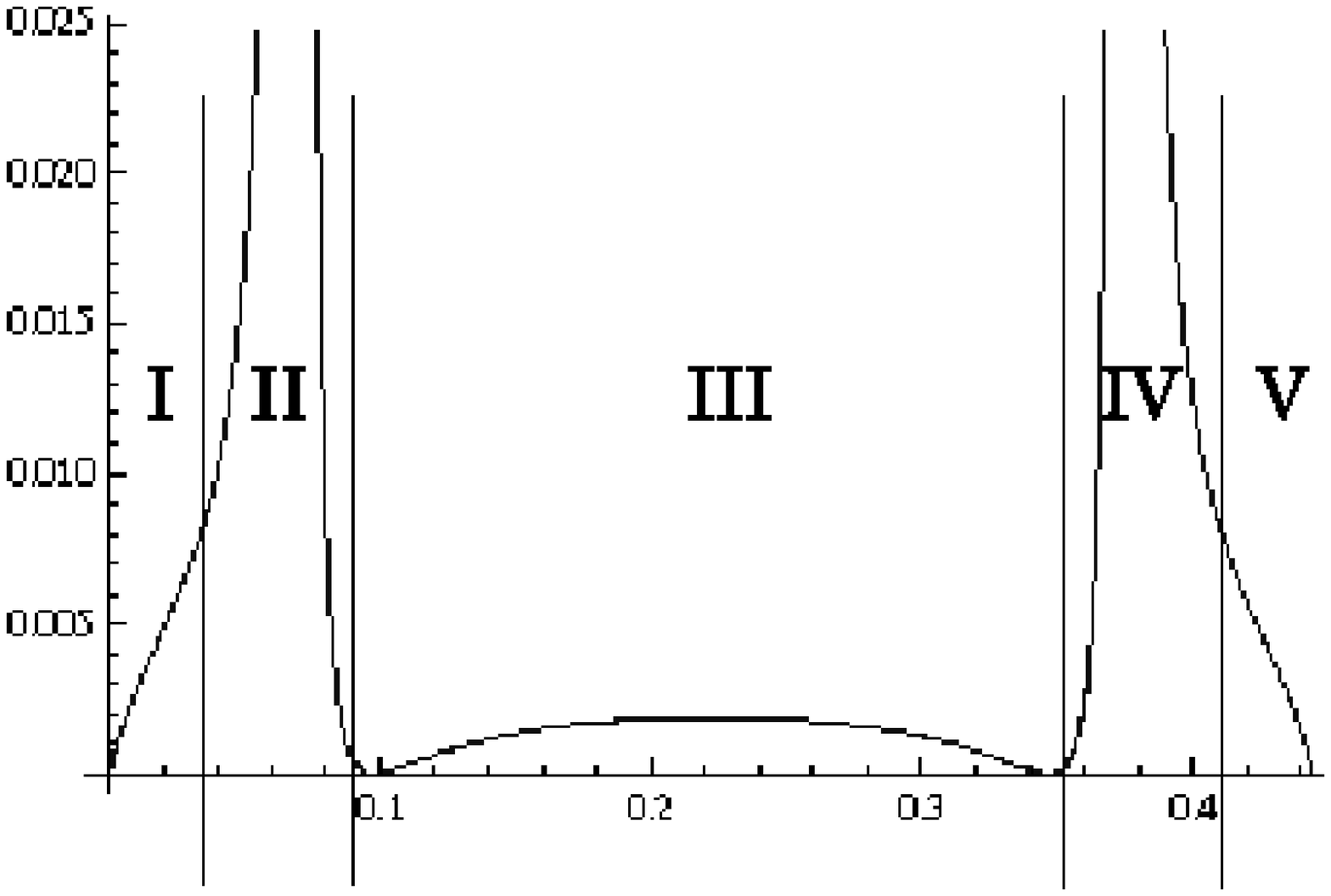}} }
\end{figure}
\small{Fig. 3. The frequency distribution  $dW^{(2\eta)}_{3s;1s}$ for the total two-photon transition $3s\rightarrow 1s+2\gamma$ including cascade and "pure" two-photon transitions as functions of the frequency (in a.u.). 
The transition rate divided by $\alpha^6$ ($\alpha$ is the fine structure constant) is plotted versus the frequency within the interval $[0,\omega_0]$. The boundaries for the frequency intervals {\bf I}-{\bf V} are also indicated
as vertical lines.
}

{
\begin{figure}[ht]
\centerline{\mbox{\epsfxsize=100pt \epsffile{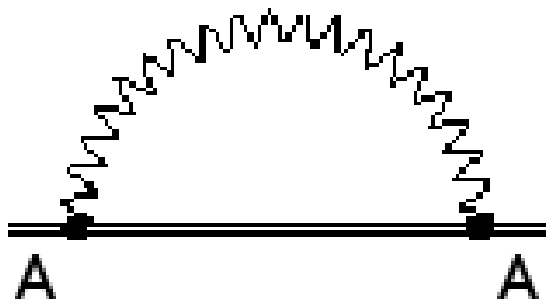}} }
\end{figure}
\small{Fig. 4. The Feynman graph corresponding to the one-loop electron self-energy. The imaginary part of the contribution of this graph to the energy level shift $\Delta E_A$ represents the one-photon width $\Gamma_A$ of the level $A$. The wavy line denotes here the virtual photon.
}}
\begin{table}[h]
\caption{
Branching ratios and transition rates (in $s^{-1}$) for the different decay channels for the decay probability of the $3s$ level with different frequency interval size ($l$).}
\small{
\begin{tabular}{| l | c | c | c | c | c | c | c | c |}
\hline \hline
$l $& $10^4$ & $10^{5}$&$2.5\cdot 10^5$& $5\cdot 10^5$ & $10^6$ & $1.5\cdot 10^6 $ &$4.53\cdot 10^6$& $1.00256\cdot 10^7$\\ \hline
$b^{(\rm{pure} 2\gamma)}$& $3.2003\cdot 10^{-5}$ & $3.5091\cdot 10^{-6}$&$1.6270\cdot 10^{-6}$&$1.0239\cdot 10^{-6}$ & $7.6765\cdot 10^{-7}$ & $ 7.2201\cdot 10^{-7}$ & $9.1487\cdot 10^{-6}$ &$1.2567\cdot 10^{-6}$\\ \hline
$W^{(\rm{pure} 2\gamma)}_{I}$ & $ 53.054$ & $ 7.0547$ &$ 3.5743$& $ 2.1898$ & $ 1.27737$ &$0.85130 $ & $2.4979\cdot 10^{-6} $& $0$\\ \hline
$W^{(\rm{pure} 2\gamma)}_{II}$ & $ 0.006247$ & $ 0.06247$ &$ 0.15614$& $ 0.31201$ & $ 0.62183$ &$0.92718 $ & $ 2.4666$& $ 3.9810$\\ \hline
$W^{(\rm{pure} 2\gamma)}_{III}$ & $ 95.536$ & $ 7.8778$ &$ 2.7928$& $ 1.4517$ & $ 1.0457$ &$1.0031 $ & $ 0.86005$& $0$\\ \hline
$W^{(\rm{pure} 2\gamma)}_{IV}$ & $ 0.006185$ & $ 0.061847$ &$0.15458$& $ 0.30890$ & $0.61569$ &$0.91813 $ & $ 2.4523$& $ 3.9575$\\ \hline
$W^{(\rm{pure} 2\gamma)}_{V}$ & $ 53.561$ & $ 7.1101$ &$ 3.5999$& $ 2.2056$ & $ 1.2886$ &$0.861254 $ & $ 3.1665\cdot 10^{-4}$& $0$\\ \hline
$W^{(\rm{pure} 2\gamma)} $ & $202.16$ & $22.167$ &$10.278$& $6.4680$ & $4.8492$ &$4.5609$ & $5.7792$& $7.9385$\\ \hline 
$b^{(\rm{inter})} $ & $-1.4342\cdot 10^{-9}$ & $-1.4343\cdot 10^{-8}$& $-3.5852\cdot 10^{-8}$& $-7.1665\cdot 10^{-8}$ & $-1.4302\cdot10^{-7}$ &$ -2.1376\cdot 10^{-7}$ & $-6.0829\cdot 10^{-7}$ & $-1.0459\cdot 10^{-6}$\\ \hline
$W^{(\rm{inter})} $ & $-0.0090599$ & $-0.090602$& $-0.22647$& $-0.45270$ & $-0.90346$ &$-1.3503$ & $-3.8426$ &$-6.6067$\\
\hline \hline
\end{tabular}
}
\end{table}
\end{document}